[1994/12/01]
\documentclass[fdp,a4paper,fleqn]{w-art}
\usepackage{times}
\usepackage{amsmath}
\usepackage{w-thm}
\usepackage{bm}
\usepackage{bbm}                                          \usepackage[]{graphicx}

\chardef\bslash=`\\ 
\newcommand{\ntt}{\normalfont\ttfamily}

\newcommand{\pkg}[1]{{\protect\ntt#1}}

\newcommand{\eval}[2][\right]{\relax
  \ifx#1\right\relax \left.\fi#2#1\rvert}

\newcommand{\nc}{\newcommand}
\nc{\hc}{\hbox {H.c.}} 
\nc{\noi}{\noindent}
\nc{\barx}{\bar{x}}
\nc{\pbarn}{\;\hbox {pb}}
\nc{\fbarn}{\;\hbox {fb}}
\nc{\lsp}{\;\;\;\;\;}
\nc{\Lsp}{\;\;\;\;\;\;\;\;\;\;}  
\nc{\LLsp}{\lspace \lspace}
\nc{\lra}{\longrightarrow}
\nc{\beq}{\begin{equation}}  \nc{\eeq}{\end{equation}}
\nc{\bea}{\begin{eqnarray}}  \nc{\eea}{\end{eqnarray}}
\nc{\baa}{\begin{array}}     \nc{\eaa}{\end{array}}
\nc{\bit}{\begin{itemize}}   \nc{\eit}{\end{itemize}}
\nc{\ben}{\begin{enumerate}} \nc{\een}{\end{enumerate}}
\nc{\bce}{\begin{center}}    \nc{\ece}{\end{center}}
\nc{\bpm}{\begin{pmatrix}}   \nc{\epm}{\end{pmatrix}}
\nc{\bvt}{\begin{verbatim}}  \nc{\evt}{\end{verbatim}}

\def\gesim{\,{\raise-3pt\hbox{$\sim$}}\!\!\!\!\!{\raise2pt\hbox{$>$}}\,}
\def\lesim{\,{\raise-3pt\hbox{$\sim$}}\!\!\!\!\!{\raise2pt\hbox{$<$}}\,}

\def\gev{\;\hbox{GeV}}
\def\tev{\;\hbox{TeV}}

\def\lsp{\qquad}

\def\gsim{\gesim}
\def\hc{\hbox{H.c.}}
\def\vev{vacuum expectation value}

\def\zBB{{\mathbbm Z}}

\nc{\lam}{\lambda}
\nc{\Lam}{\Lambda}
\nc{\Lams}{\Lambda^2}
\nc{\mws}{m_W^2}
\nc{\mzs}{m_Z^2}
\nc{\mts}{m_t^2}
\nc{\mh}{m_h}
\nc{\mhs}{m_h^2}
\nc{\mvp}{m_\vp}
\nc{\mvps}{m_\vp^2}
\nc{\mw}{m_W}
\nc{\mz}{m_Z}
\nc{\mt}{m_t}
\nc{\mH}{m_{H^\pm}}
\nc{\mA}{m_A}
\nc{\mS}{m_S}
\nc{\vp}{\varphi}
\nc{\mpl}{m_{\rm Pl}}
\nc{\sbb}{s_\beta}
\nc{\cbb}{c_\beta}
\nc{\sba}{s_{\beta-\alpha}}
\nc{\cba}{c_{\beta-\alpha}}
\nc{\stb}{s_{2\beta}}
\nc{\ctb}{c_{2\beta}}
\nc{\mb}{m_b}
\nc{\mbs}{m_b^2}
\nc{\tgb}{\tan\beta}
\nc{\tgbs}{\tan^2\beta}
\nc{\ctbs}{\cot^2\beta}

\nc{\lamp}{\lambda_H}
\nc{\lamvp}{\lambda_\varphi}
\nc{\lamx}{\lambda_x}
\nc{\xf}{x_f}

\nc{\co}{{\bf ??? }}

\renewcommand{\Im}{{\rm Im\thinspace}}

\begin{document}
\Volume{XX}
\Issue{1}
\Month{1}
\Year{2003}
\pagespan{1}{}
\Receiveddate{XX}
\Reviseddate{XX}
\Accepteddate{XX}
\Dateposted{XX}




\title[Short Title]{Natural Two-Higgs-Doublet Model}


\author[Bohdan Grzadkowski]{Bohdan Grzadkowski\footnote{Corresponding author:
    e-mail: {\sf bohdan.grzadkowski@fuw.edu.pl}, Phone: +48\,225\,532\,259, Fax:
    +48\,226\,219\,475}\inst{1}} \address[\inst{1}]{Institute of Theoretical Physics, Faculty of Physics, University of Warsaw, Ho\.za 69, PL-00-681 Warsaw, Poland}

\author[Per Osland]{Per Osland\inst{2}}
\address[\inst{2}]{Department of Physics and Technology, University of Bergen,
Postboks 7803, N-5020 Bergen, Norway}


\thanks{}

\begin{abstract}
  We show that the Two-Higgs-Doublet Model (2HDM) constrained by the 
  two-loop-order requirement of
  cancellation of quadratic divergences 
  is consistent with the existing experimental constraints. The model 
  allows to ameliorate the little hierarchy problem by suppressing the 
  quadratic corrections to scalar masses and lifting the mass of the lightest Higgs
  boson. A strong source of CP violation emerges from the scalar potential.  
  The cutoff originating from the naturality arguments
  is shifted from $\sim 0.6\tev$ in the Standard Model 
  to $\gsim 6 \tev$ in the 2HDM, depending on the mass of the
  lightest scalar. 
\end{abstract}

\maketitle





\renewcommand{\leftmark}
{\textsc{WILEY-VCH Verlag Berlin GmbH}: Sample paper for the \pkg{w-art} class}


\section{Introduction}
We are going to discuss an extension of the Standard Model (SM) that 
is free of quadratic divergences 
up to the leading order at the two-loop level of the perturbation expansion.  
The quadratic divergences were first discussed within the SM by
Veltman~\cite{Veltman:1980mj}, who, adopting dimensional
reduction~\cite{Siegel:1979wq}, found the following quadratically divergent
one-loop contribution to the Higgs boson ($h$) mass 
$\delta^{\rm (SM)} \mhs = \Lams/(\pi^2 v^2)(\frac32
\mts-(6\mws+3\mzs)/8 - 3\mhs/8 )$ with $ \Lam$ being a UV cutoff 
and $v \simeq 246 \gev $ denoting the \vev\ of the scalar doublet. The
issue of quadratic divergences was then investigated adopting 
other regularization schemes (e.g.\ point
splitting~\cite{Osland:1992ay}) and also in
\cite{Einhorn:1992um} without reference to any regularization scheme.

Precision measurements within the SM imply a small Higgs-boson mass, therefore the
correction $\delta^{\rm (SM)} \mhs$ exceeds the mass itself even for small values
of $ \Lam $, e.g. for $\mh = 130 \gev$ one obtains $\delta^{\rm (SM)}
\mh^2 \simeq \mh^2$ already for $\Lam \simeq 600 \gev$. However, if one assumes that the scale of new physics is widely separated
from the electro-weak scale, then constraints that arise from
analysis of operators of dimension 6 require $\Lam \gsim$ a few TeV.
The conclusion that follows from this observation is that regardless of what physics lies beyond the SM, 
some amount of fine tuning is necessary; either one tunes to
lift the cutoff above $\Lam \simeq 600 \gev$, or one tunes when
precision observables measured at LEP are fitted.
Tuning both in corrections to the Higgs mass and in LEP physics is  
also an acceptable alternative which we are going to explore below.
In other words, we will look for new physics in the TeV range which will allow to
lift the cutoff implied by quadratic corrections to $\mhs$ to the
multi-TeV range {\it and} which will be consistent with all the
experimental constraints---both require some amount of tuning.  It should be realized 
that within the SM the requirement $\delta^{\rm (SM)} \mhs = 0$
implies an unrealistically large  Higgs boson mass $\mh \simeq
310~\gev$.

Here we will argue that within the
Two-Higgs-Doublet Model (2HDM) one can 
soften the little hierarchy problem both by suppressing quadratic corrections
to scalar masses {\it  and} it allows to lift the central value for the 
lightest Higgs mass up to a value which is well above the LEP limit.

\section{ The Two-Higgs-Doublet Model}
\label{non-IDM}

In order to accommodate CP violation we consider here a 2HDM
with softly broken $\zBB_2$ symmetry which acts as $\Phi_1\to -\Phi_1$ and $u_R\to -u_R$ (all other fields are neutral). The scalar potential then reads
\begin{eqnarray}
V(\phi_1,\phi_2) &=&  -\frac12 \left\{m_{11}^2\phi_1^\dagger\phi_1 
+ m_{22}^2\phi_2^\dagger\phi_2 + \left[m_{12}^2 \phi_1^\dagger\phi_2 
+ \hc \right]\right\}
 + \frac12 \lam_1 (\phi_1^\dagger\phi_1)^2 
+ \frac12 \lam_2 (\phi_2^\dagger\phi_2)^2 
\nonumber  \\
&& 
+ \lambda_3(\phi_1^\dagger\phi_1)(\phi_2^\dagger\phi_2) 
+ \lambda_4(\phi_1^\dagger\phi_2)(\phi_2^\dagger\phi_1) 
+ \frac12\left[\lambda_5(\phi_1^\dagger\phi_2)^2 + \hc\right] 
\label{2HDMpot}
\end{eqnarray}
The minimum of the potential is achieved at $\langle \phi_1^0 \rangle = v_1/\sqrt{2}$ and 
$\langle \phi_2^0 \rangle = v_2/\sqrt{2}$.
We assume that $\phi_1$ and $\phi_2$ couple to down- and up-type
quarks, respectively (the so-called 2HDM II).  

\subsection{Quadratic divergences}
\label{one-loop}

At the one-loop level the cancellation of quadratic divergences for the 2-point scalar Green's functions 
at zero external momenta ($\Gamma_i$, $i=1,2$) in the 2HDM 
type II model implies~\cite{Newton:1993xc} 
\begin{eqnarray}
\Gamma_1\equiv \frac32 \mw^2 + \frac34 \mz^2 
+ \frac{v^2}{2}\left( \frac32 \lam_1 + \lam_3 + \frac12 \lam_4 \right) 
- 3 \frac{\mb^2}{\cbb^2} = 0,
\label{qdcon1_mod2}\\
\Gamma_2\equiv\frac32 \mw^2 + \frac34 \mz^2 
+ \frac{v^2}{2}\left( \frac32 \lam_2 + \lam_3 + \frac12 \lam_4 \right) 
-3 \frac{\mt^2}{\sbb^2} = 0,
\label{qdcon2_mod2}
\end{eqnarray} 
where $v^2\equiv v_1^2+v_2^2$, $\tan\beta\equiv v_2/v_1$ and we use the
notation: $s_\beta \equiv \sin\beta$ and $c_\beta\equiv \cos\beta$. 
In the type II model
the mixed, $\phi_1-\phi_2$, Green's function is not quadratically divergent.
Some phenomenological consequences of the cancellation were
discussed already in \cite{Ma:2001sj}.  

As shown in \cite{ElKaffas:2007rq},  the quartic couplings $\lambda_i$ can be
expressed in terms of the mass parameters and elements of the rotation matrix 
needed for diagonalization of the scalar masses. 
So, for a given choice of $\alpha_i$'s, the
squared neutral-Higgs masses $M_{1}^2$, $M_{2}^2$ and $M_3^2$ can be
determined from the cancellation conditions
(\ref{qdcon1_mod2})--(\ref{qdcon2_mod2}) in terms of $\tgb$, $\mu^2$
and $M_{H^\pm}^2$.
It is worth noticing that scalar masses resulting from a scan over $\alpha_i$, $M_{H^\pm}$ and 
$\tgb$ exhibit a striking mass
degeneracy in the case of
large $\tan\beta$: $M_1 \simeq M_2 \simeq M_3
\simeq \mu^2+4\mb^2$. 

At the two-loop level the leading contributions to quadratic divergences are
of the form of $ \Lam^2 \ln \Lam$. They could be determined adopting a strategy 
noticed by Einhorn and Jones~\cite{Einhorn:1992um}, so that the cancellation 
conditions for quadratic divergences up to the leading two-loop order read:
\beq
\Gamma_1+\delta \Gamma_1=0 \lsp {\rm and} \lsp \Gamma_2+\delta \Gamma_2=0 
\label{2-loop-con}
\eeq
with
\bea
\delta \Gamma_1 &=& \frac{v^2}{8} [
9 g_2 \beta_{g_2} + 3 g_1 \beta_{g_1} + 6\beta_{\lambda_1} + 4 \beta_{\lambda_3} + 2 \beta_{\lambda_4}]\ln\left(\frac{\Lambda}{\bar\mu}\right)\\
\delta \Gamma_2 &=& \frac{v^2}{8} [
9 g_2 \beta_{g_2} + 3 g_1 \beta_{g_1} + 6\beta_{\lambda_2} + 4 \beta_{\lambda_3} + 2 \beta_{\lambda_4}
-24 g_t \beta_{g_t}]\ln\left(\frac{\Lambda}{\bar\mu}\right)
\eea
where $\beta$'s are the appropriate beta functions while $\bar \mu$ is the renormalization scale.
We will be solving the conditions (\ref{2-loop-con}) for the scalar masses $M_i^2$ for a given set of $\alpha_i$'s, $\tgb$, $\mu^2$ and $M_{H^\pm}^2$. For the renormalization scale
we will adopt $v$, so $\bar\mu=v$. Those masses together with the corresponding coupling constants, will be used to find predictions of the model for various observables.

\begin{figure}[t]
\centering
\includegraphics[width=4.5cm, height=4cm]{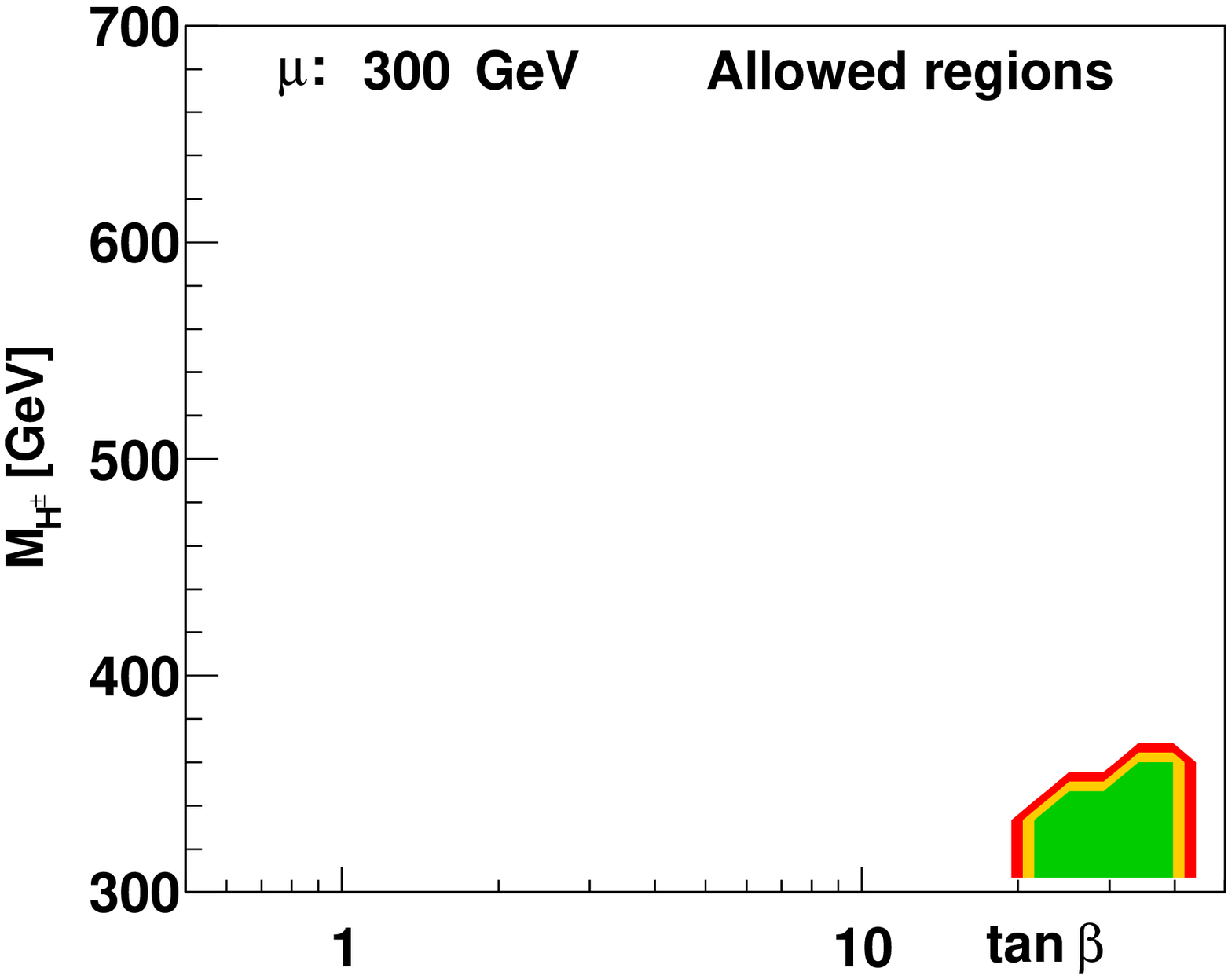}
\includegraphics[width=4.5cm, height=4cm]{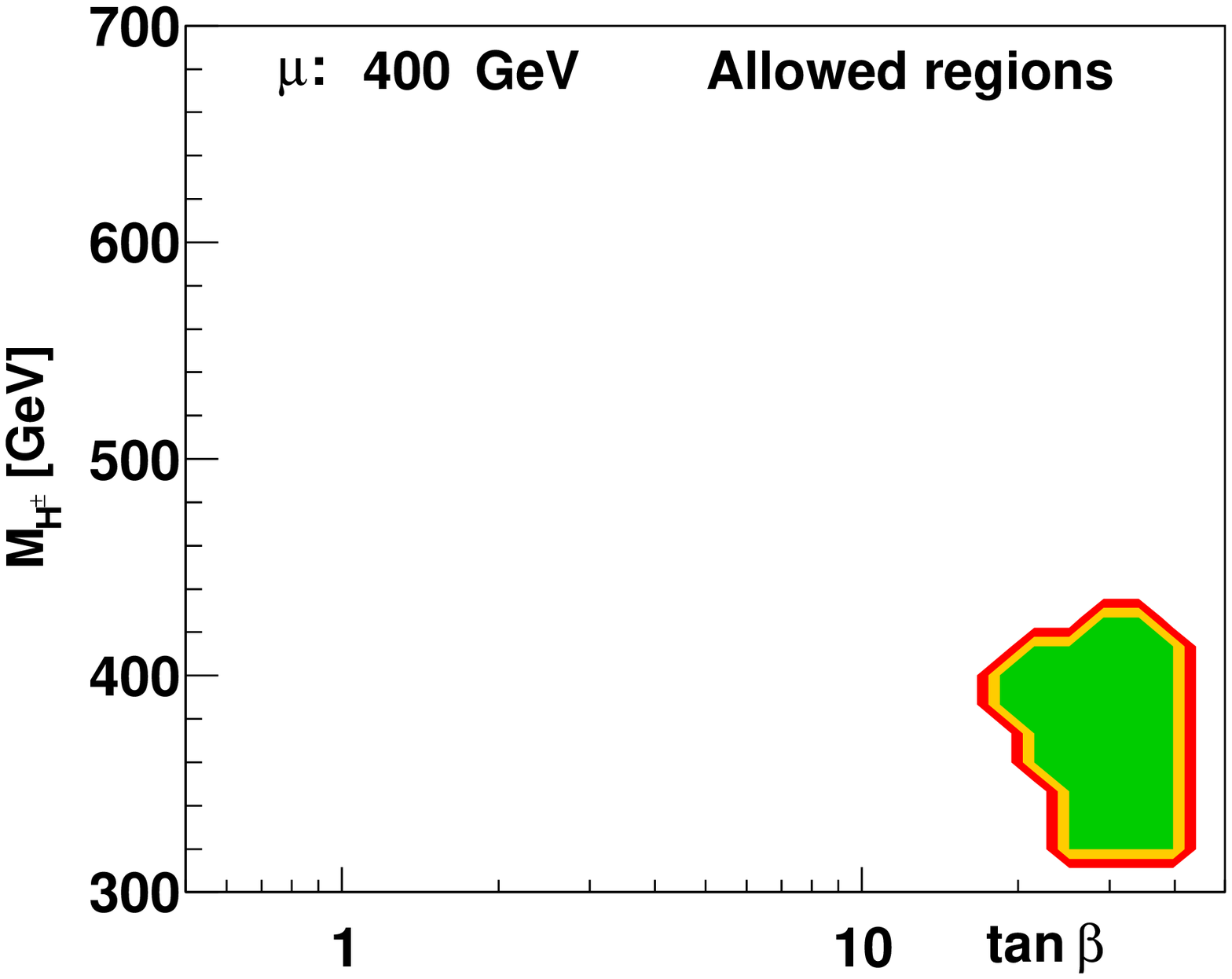}
\includegraphics[width=4.5cm, height=4cm]{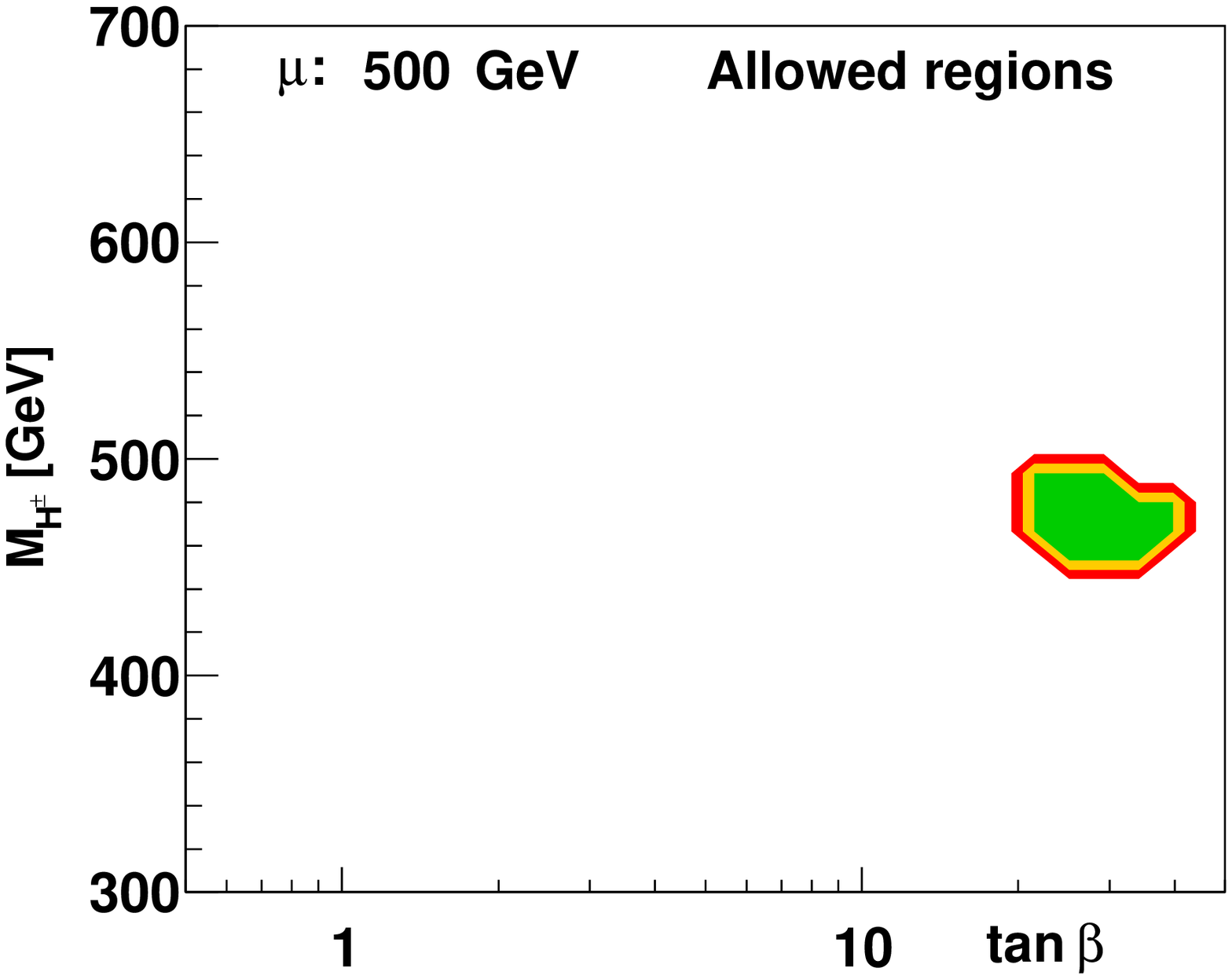}
\caption{
  Two-loop allowed regions in
  the $\tan\beta$--$M_{H^\pm}$ plane, for $\Lam=4.5\tev$, for $\mu=300, 400, 500\gev$ 
  (as indicated).  Red: positivity
  is satisfied; yellow: positivity and unitarity both satisfied;
  green: also experimental constraints satisfied at the 95\% C.L., as
  specified in the text. }
\label{Fig:allowed-4500-300-400-500}
\end{figure}

\subsection{Allowed regions}
\label{sec:allowed}

In order to find phenomenologically acceptable regions in the 
parameter space we impose the following experimental constraints:
the oblique parameters $T$ and $S$, $B_0-\bar{B}_0$ mixing, $B\to X_s \gamma$,
$B\to \tau \bar\nu_\tau X$, $B\to D\tau \bar\nu_\tau$, LEP2 Higgs-boson non-discovery,
$R_b$, the muon anomalous magnetic moment and the electron electric dipole moment
(for details concerning the experimental constraints, see
refs.~\cite{Grzadkowski:2009bt,ElKaffas:2007rq,WahabElKaffas:2007xd}).
Subject to all these constraints, we find allowed solutions of (\ref{2-loop-con}). 
For instance, imposing the positivity constraints 
we find allowed regions in the $\tan\beta$--$M_{H^\pm}$ plane as 
illustrated by the red domains in the $\tan\beta$--$M_{H^\pm}$ plane, see
Fig.~\ref{Fig:allowed-4500-300-400-500} 
for fixed values of $\mu$. The allowed regions
were obtained scanning over the mixing angles $\alpha_i$ and solving
the two-loop cancellation conditions (\ref{2-loop-con}). Imposing
also unitarity in the Higgs-Higgs-scattering sector 
\cite{Kanemura:1993hm,Akeroyd:2000wc,Ginzburg:2003fe}, the allowed regions are only slightly
reduced (yellow regions). Requiring that also experimental constraints 
are satisfied the green regions are obtained. 

\begin{figure}[ht]
\centering
\includegraphics[width=10cm, height=4cm]{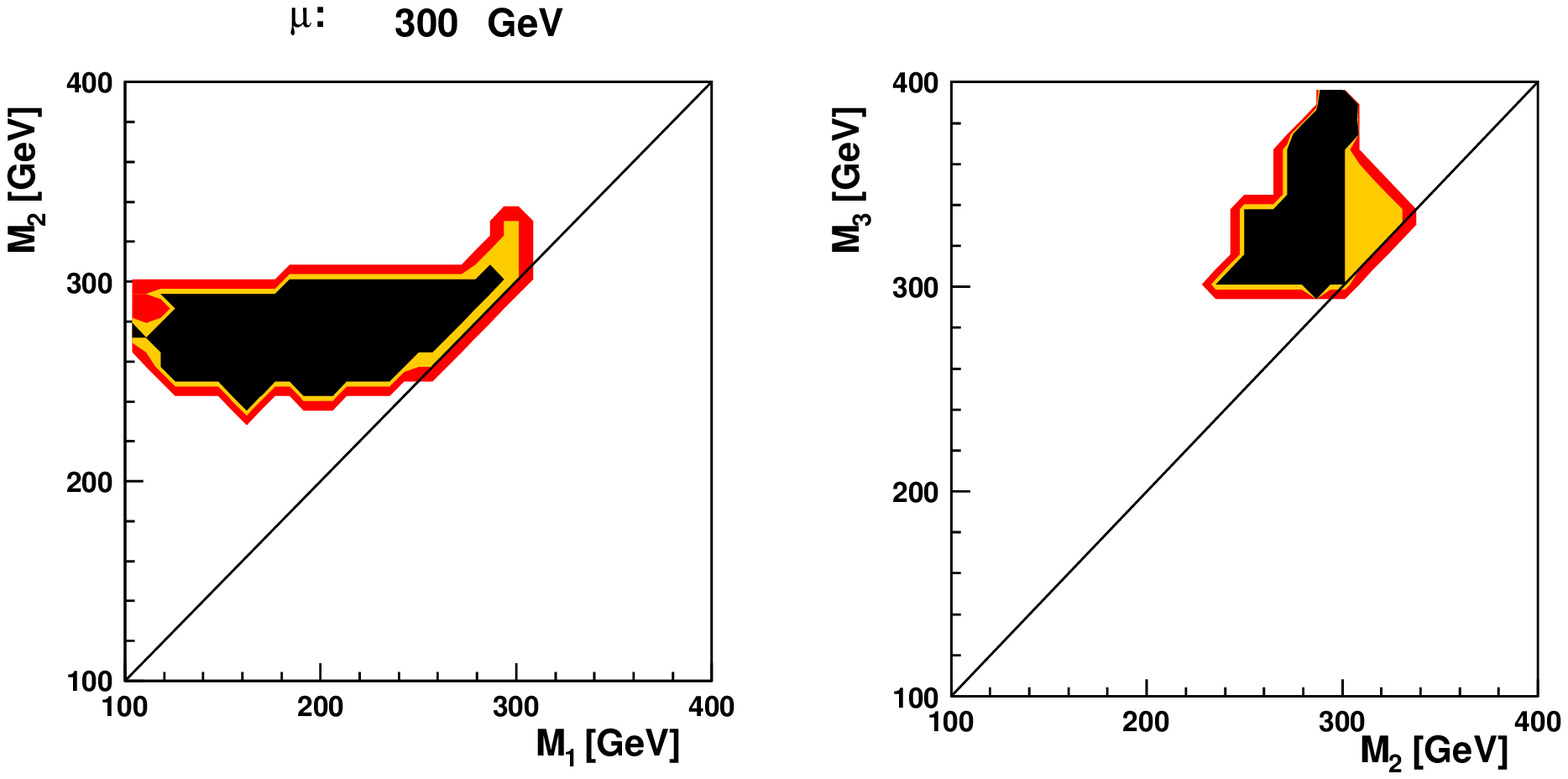}
\includegraphics[width=10cm, height=4cm]{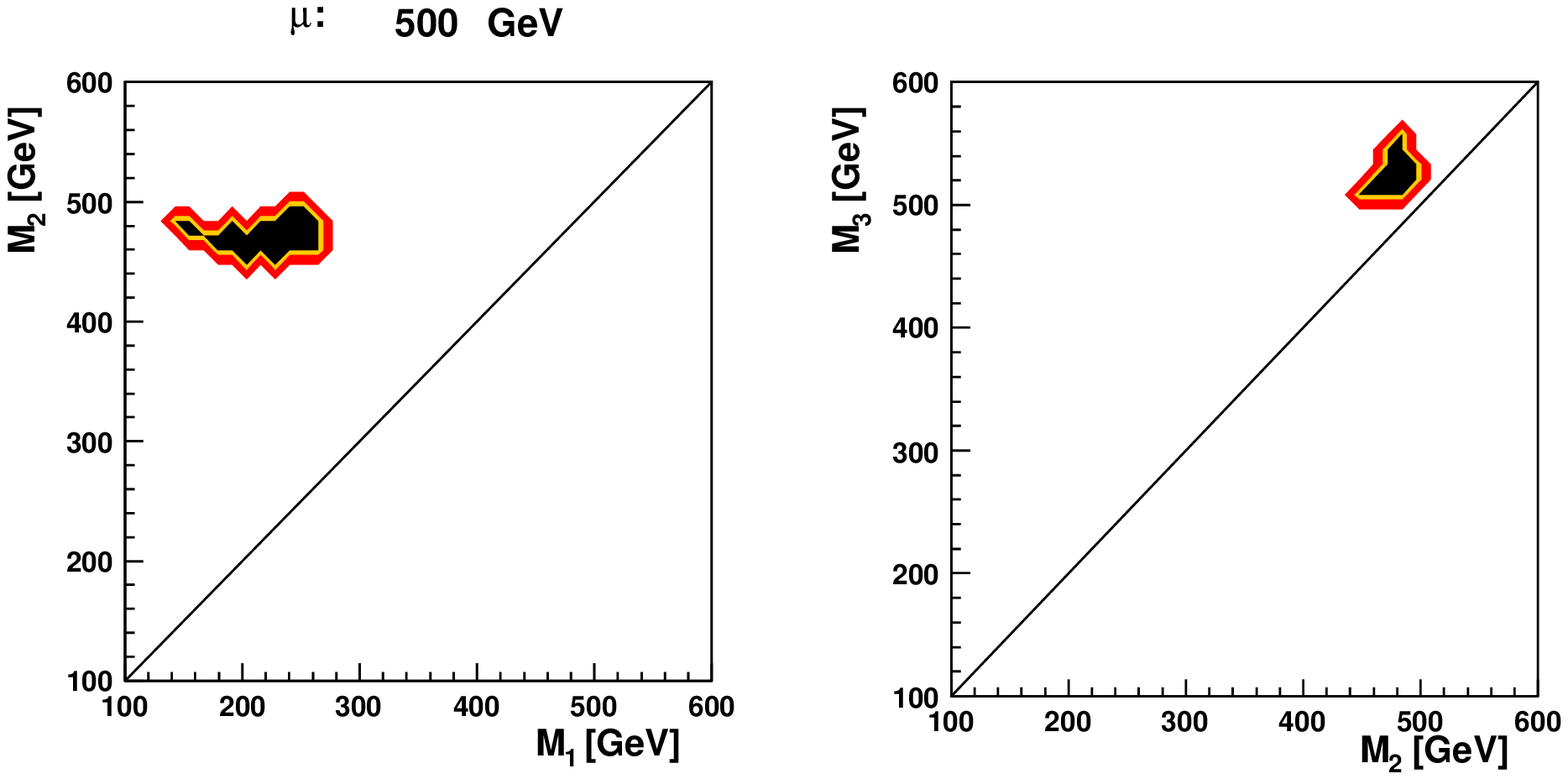}
\caption{
  Two-loop distributions of allowed masses $M_2$ vs $M_1$ (left panels)
  and $M_3$ vs $M_2$ (right) for $\Lam=4.5\tev$, resulting from a scan over the full
  range of $\alpha_i$, $\tan\beta \in (0.5,50)$ and $M_{H^\pm} \in
  (300,700)\gev$, for $\mu=300, 500~{\rm GeV}$. Red: Positivity
  is satisfied; yellow: positivity and unitarity both satisfied;
  green: also experimental constraints satisfied at the 95\% C.L., as
  specified in the text. }
\label{Fig:2-loop-masses-4500}
\end{figure}

For parameters that are consistent with unitarity, positivity, experimental
constraints and the two-loop cancellation conditions (\ref{2-loop-con}), we show in Fig.~\ref{Fig:2-loop-masses-4500} scalar masses resulting from a scan over $\alpha_i$, 
$M_{H^\pm}$ and $\tgb$. 

\subsection{CP violation}
\label{cpv}

CP violation could be parametrized adopting the 
$U(2)$-invariants introduced by Lavoura and Silva \cite{Lavoura:1994fv},
\cite{Branco:2005em}. It is convenient to use the invariants $J_1$, $J_2$ and $J_3$ 
defined in \cite{Gunion:2005ja}.
The Higgs sector is
CP-conserving if and only if $\Im J_i=0$ for all $i$ \cite{Gunion:2005ja}. 
The invariants calculated in the basis adopted
here could be found in \cite{Grzadkowski:2009bt}.
As we have noticed earlier large $\tan\beta$ implies approximate
degeneracy of scalar masses. That could jeopardize the CP
violation in the potential since it is well known that the exact degeneracy $M_1=M_2=M_3$
results in vanishing invariants $\Im J_i$ and no CP violation (exact
degeneracy implies $\Im \lambda_5=0$). Adopting the one-loop conditions
(\ref{qdcon1_mod2})--(\ref{qdcon2_mod2}) one easily finds that
$\lambda_1-\lambda_2=4(\mb^2/\cbb^2-\mt^2/\sbb^2)/v^2$, which implies
\begin{equation}
\Im J_1 = 4\, \Im \lambda_5 (\cbb^2\mt^2-\sbb^2 \mb^2)/v^2
=-4\, \Im \lambda_5(\mb/v)^2 + 
{\cal{O}}(\Im \lambda_5/\tgb^2)
\label{imj1}
\end{equation}
If
$\tgb$ is large then $\Im J_1$ is reduced not only by $\Im
\lambda_5 \simeq 0$ (as caused by $M_1\simeq M_2\simeq M_3$) but also
by the factor $(\mb^2/v^2)$, as implied by the cancellation conditions
(\ref{qdcon1_mod2})--(\ref{qdcon2_mod2}). The same suppression factor
appears for $\Im J_3$. The case of $\Im J_2$ is more involved, however
when $\mb^2/v^2$ is neglected all the invariants
have the same 
behavior for large $\tgb$:
\begin{equation}
\Im J_i \sim \Im \lambda_5/\tgbs
\end{equation}
Those conclusions apply also
at the two-loop level. It turns out that at high values of $\tan\beta$ these 
invariants are of the
order of $10^{-3}$, in qualitative agreement with the discussion
above. It is worth emphasizing 
that the corresponding invariant in the SM; $\Im Q = \Im (V_{ud} V_{cb} V_{ub}^\star V_{cd}^\star)$ is of the order of 
$\sim 2\times 10^{-5} \sin \delta_{KM}$ ($V_{ij}$ and $\delta_{KM}$ are elements of the CKM matrix and CP-violating phase, respectively). Therefore the model considered here offers much stronger CP violation than in the SM.

\section{Summary}
\label{sum}
It has been shown that within the Two-Higgs-Doublet
Model type II there exist regions of the parameter space where
the quadratic divergences in scalar boson masses are suppressed. 
The little hierarchy problem is therefore ameliorated. 
The UV cutoff could be shifted up to $\sim 6\tev$.
CP-violation emerging from the scalar potential turns out to be much stronger than
within the SM.


\begin{acknowledgement}
This work is supported in part by the Ministry of Science and Higher
Education (Poland) as research project N~N202~006334 (2008-11). 
The research of P.~O. has been supported by the Research Council of
Norway.
\end{acknowledgement}


\begin{thebibliography}{[1]}

\bibitem{Veltman:1980mj}
  M.~J.~G.~Veltman, Acta Phys.\ Polon B \textbf{12}, 437 (1981).

\bibitem{Siegel:1979wq}
  W.~Siegel,  Phys.\ Lett.\  B \textbf{84}, 193 (1979);
  D.~M.~Capper, D.~R.~T.~Jones and P.~van Nieuwenhuizen,
  Nucl.\ Phys.\  B \textbf{167}, 479 (1980).

\bibitem{Osland:1992ay}
  P.~Osland and T.~T.~Wu,
  Z.\ Phys.\  C \textbf{55}, 569 (1992);
  Z.\ Phys.\  C \textbf{55}, 585 (1992).

\bibitem{Einhorn:1992um}
  M.~B.~Einhorn and D.~R.~T.~Jones,
  Phys.\ Rev.\  D \textbf{46}, 5206 (1992).
  
\bibitem{Newton:1993xc}
  C.~Newton and T.~T.~Wu,
  Z.\ Phys. C \textbf{62} 253 (1994).

\bibitem{Ma:2001sj}
  E.~Ma,
  Int.\ J.\ Mod.\ Phys.\  A \textbf{16}, 3099 (2001)
  [arXiv:hep-ph/0101355].

\bibitem{ElKaffas:2007rq}
  A.~W.~El Kaffas, P.~Osland and O.~M.~Ogreid,
  Nonlin.\ Phenom.\ Complex Syst.\  \textbf{10}, 347 (2007)
  [arXiv:hep-ph/0702097].
  
\bibitem{Grzadkowski:2009bt}
  B.~Grzadkowski, O.~M.~Ogreid and P.~Osland,
  Phys.\ Rev.\  D \textbf{80}, 055013 (2009)
  [arXiv:0904.2173 [hep-ph]].

\bibitem{WahabElKaffas:2007xd}
  A.~W.~El Kaffas, P.~Osland and O.~M.~Ogreid,
  Phys.\ Rev.\  D \textbf{76}, 095001 (2007)
  [arXiv:0706.2997 [hep-ph]].

\bibitem{Kanemura:1993hm}
  S.~Kanemura, T.~Kubota and E.~Takasugi,
  Phys.\ Lett.\ B \textbf{313}, 155 (1993)
  [arXiv:hep-ph/9303263].
  
\bibitem{Akeroyd:2000wc}
  A.~G.~Akeroyd, A.~Arhrib and E.~M.~Naimi,
  Phys.\ Lett.\ B \textbf{490}, 119 (2000)
  [arXiv:hep-ph/0006035];
  A.~Arhrib,  arXiv:hep-ph/0012353.
 
\bibitem{Ginzburg:2003fe}
  I.~F.~Ginzburg and I.~P.~Ivanov, arXiv:hep-ph/0312374;
  Phys.\ Rev.\ D \textbf{72}, 115010 (2005)
  [arXiv:hep-ph/0508020].

\bibitem{Lavoura:1994fv}
  L.~Lavoura and J.~P.~Silva,
  Phys.\ Rev.\  D \textbf{50}, 4619 (1994)
  [arXiv:hep-ph/9404276].

\bibitem{Branco:2005em}
  G.~C.~Branco, M.~N.~Rebelo and J.~I.~Silva-Marcos,
  Phys.\ Lett.\  B \textbf{614}, 187 (2005)
  [arXiv:hep-ph/0502118].

\bibitem{Gunion:2005ja}
  J.~F.~Gunion and H.~E.~Haber,
  Phys.\ Rev.\  D \textbf{72} (2005) 095002
  [arXiv:hep-ph/0506227].


\end{thebibliography}
\end{document}